\documentclass[aps, preprint,amsfonts, amssymb,groupedaddress, floatfix]{revtex4}
\usepackage[english]{babel}
\usepackage{epsfig}
\usepackage{amsfonts, amssymb, amsmath}
\usepackage{times, mathptm}
\usepackage{psfrag}
\newcommand{\Tr}{\mathop{\rm Tr}}
\newcommand{\sign}{\mathop{\rm sign}}
\newcommand{\Erf}{\mathop{\rm Erf}}
\newcommand{\arcsh}{\mathop{\rm arcsh}}
\begin{document}

\title{Coherent transport in Josephson-junction rhombi chain chain with quenched disorder}

\author{Ivan V. Protopopov and Mikhail V. Feigel'man}

\affiliation{L.D.Landau Institute for Theoretical Physics,Kosygin
str.2,  Moscow, 119334, Russia}

\begin{abstract}
We consider  a chain of  Josephson-junction rhombi (proposed
originally in~\cite{Doucot}) in quantum regime. In a regular chain with no disorder in the maximally frustrated case
when magnetic flux through each rhombi $\Phi_r$ is equal to one half of superconductive flux quantum $\Phi_0$,
Josephson current is due to correlated transport of {\em pairs of Cooper pairs}, i.e. charge is quantized in units of $4e$.
Sufficiently strong deviation $\delta\Phi \equiv |\Phi_r-\Phi_0/2| > \delta\Phi^c$ from the
maximally frustrated point brings the system back to usual $2e$-quantized supercurrent.
For a regular chain $\delta\Phi^c$ was calculated in~\cite{my}. Here we present detailed analysis of the effect
of quenched disorder (random stray charges and random fluxes piercing rhombi) on the pairing effect.
\end{abstract}

\maketitle
\section{Introduction}
Pairing of Cooper pairs  in frustrated Josephson junction arrays
was theoretically proposed
recently~\cite{Doucot,IoffeFeigelman02,Doucot03,Doucot04} in the
search of topologically protected nontrivial quantum liquid
states. The simplest system where such a phenomenon could be
observed was proposed by Dou\c{c}ot and Vidal in~\cite{Doucot}. It
consists of a chain of rhombi (each of them being small ring of 4
superconductive islands connected by 4 Josephson junctions) placed
into transverse magnetic field. cf. Fig. ~1. It was shown
in~\cite{Doucot} that in the fully frustrated case (i.e. magnetic
flux through each rhombus $\Phi_r = \frac12\Phi_0 = \frac{hc}{4e}
$) usual tunnelling of Cooper pairs along the chain is blocked due
to destructive interference of tunneling going through two paths
within the same rhombus, while  correlated 2-Cooper-pair transport
survives. Sufficiently strong deviation $ \delta\Phi \equiv
|\Phi_r-\Phi_0/2| > \delta\Phi^c$ from the maximally frustrated
point brings the system back to usual $2e$-quantized supercurrent.

In ref. \cite{my} rhombi chain was studied for the experimentally
relevant situation when its Coulomb energy is determined by
capacitance of junctions and not by capacitance of superconductive
islands themselves. Expression for the critical deflection
$\delta\Phi^c$ was derived. However, ref. \cite{my} dealt with
regular chain with no disorder. In any real system two intrinsic
sources of disorder are always present: a) some weak randomness of
fluxes $\Phi_r^n$ penetrating different rhombi (due to unavoidable
differences in their areas), and b) random stray charges $q_n$
which produce, due to Aharonov-Casher effect, some random phase
factors  to the phase slip tunnelling amplitudes leading to
suppression of quantum fluctuations in the chain  (compare
analogous discussion in \cite{Larkin}).

In this paper we adapt method used in ref. \cite{my} for the case
of rhombi chain with quenched disorder and study influence of
random stray charges and random fluxes in rhombi on the crossover
point $\delta\Phi=\delta\Phi^c$ between $4e$- and $2e$-regimes. It
turns out that in a long chain with large $E_J/E_C$ the pairing
effect is rather stable under influence of disorder.

The rest of the paper is organized as follows: in Sec. II we
derive effective Schr\"{o}dinger equation describing rhombi chain with quenched random stray charges or random fluxes
piercing rhombi; in Sec. III we discuss influence of quenched stray charges on pairing effect, derive "phase diagram" of
the chain with fixed realization of disorder, calculate probability ${\cal P}_{4e}(E_J/E_C, N, \delta\Phi)$ to
find the chain in the regime with dominating $4e$-supercurrent and finally estimate critical deflection from the
maximally frustrated point $\delta\Phi^c$ destroying $4e$-supercurrent in a chain with stray charges; in Sec. IV
we consider modulation of the supercurrent in a clean and disordered chain by external capacitively coupled gates;
in Sec. V. we analyse influence of randomness of fluxes in rhombi on  pairing of Cooper pairs. Finally,  in Sec. VI.
we present our conclusions and suggestions.

\section{Effective Hamiltonian in presence of disorder}
\begin{widetext}
\begin{figure}
\centering
\psfrag{Phir}[c][c]{\small  $\Phi_r$}
\psfrag{Phic}[c][c]{\small $\Phi_c$}
\psfrag{N}[c][c]{\small $N$}
\psfrag{N1}[c][c]{\small $N$}
\psfrag{VG}[c][c]{\small $V_g$}
\psfrag{CG}[c][c]{\small $C_g$}
\psfrag{theta1N}[c][c]{\small  $\theta_N^{(1)}$}
\psfrag{theta2N}[cl][c]{\small $\theta_N^{(2)}$}
\psfrag{theta3N}[c][c]{\small  $\theta_N^{(3)}$}
\psfrag{theta4N}[c][c]{\small  $\theta_N^{(4)}$}
\psfrag{theta11}[bl][c]{\small  $\theta_1^{(1)}$}
\psfrag{theta21}[c][c]{\small  $\theta_1^{(2)}$}
\psfrag{theta31}[c][c]{\small  $\theta_1^{(3)}$}
\psfrag{theta41}[c][c]{\small  $\theta_1^{(4)}$}
\psfrag{theta12}[c][c]{\small  $\theta_2^{(1)}$}
\psfrag{theta22}[c][c]{\small  $\theta_2^{(2)}$}
\psfrag{theta32}[c][c]{\small  $\theta_2^{(3)}$}
\psfrag{theta42}[c][c]{\small  $\theta_2^{(4)}$}
\psfrag{n1}[c][c]{\small  $1$}
\psfrag{n2}[c][c]{\small  $2$}
\psfrag{q11}[c][c]{\small  $q_1^1$}
\psfrag{q12}[c][c]{\small  $q_1^2$}
\psfrag{q13}[c][c]{\small  $q_1^3$}
\psfrag{q21}[c][c]{\small  $q_2^1$}
\psfrag{q22}[c][c]{\small \vspace{0.2cm} $q_2^2$}
\psfrag{q23}[c][c]{\small  $q_2^3$}
\psfrag{qN2}[c][c]{\small  $q_N^2$}
\psfrag{qN3}[c][c]{\small  $q_N^3$}
\psfrag{qN1}[c][c]{\small  $q_N^1$}
\includegraphics[width=400pt]{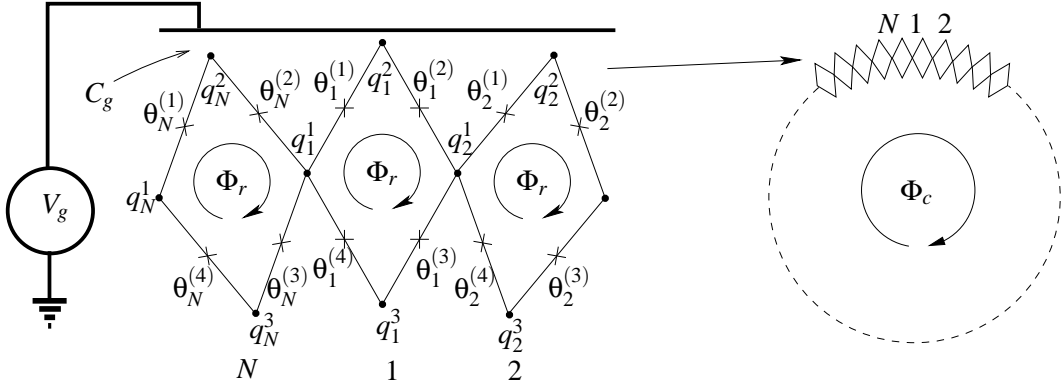}
\caption{\small The chain of rhombi with random stray charges. Also shown is an external capacitively
coupled gate (c.f. Sec. IV).}
\label{chain}
\end{figure}
\end{widetext}

We study a chain of $N$ rhombi shown in Fig.~\ref{chain}. Each
rhombi consists of four superconductive islands connected by
tunnel junctions with Josephson coupling energy $E_J=\hbar
I_c^0/2e$; charging energy $E_C$ is determined by capacitance  $C$
of junctions, $E_C = e^2/2C$ (we neglect self-capacitances of
islands which are assumed to be much smaller than $C$).  Below we
consider Josephson current along the chain of $N \gg 1$ rhombi and
assume that the chain is of the ring shape, with total magnetic
flux $\Phi_c$ inside the ring. We also denote by $\Phi_r^n$ the
flux through $n$-th rhombus and define phases $\gamma$ and
$\varphi_n$:
\begin{equation}
\gamma=2\pi \frac{\Phi_c}{\Phi_0}\,, \qquad
\varphi_n=2\pi\frac{\Phi_r^n}{\Phi_0}\,,
\end{equation}
A regular chain with no disorder (no random stray charges and all
$\varphi_n\equiv \varphi$) is described by the imaginary-time
action
\begin{equation} S_E=\int
dt\sum_{n=1}^N\sum_{m=1}^4\left\{\frac{1}{16E_C}\left(\frac{d\theta^{(m)}_n}{dt}
\right)^2-E_J\cos\theta^{(m)}_n\right\}. \label{old_basic_action}
\end{equation}
and additional conditions
\begin{equation}
\sum_{m=1}^4\theta^{(m)}_n=\varphi\,, \qquad n=1,2,..,N\,,
\label{varphi}
\end{equation}
\begin{equation}
\sum_{n=1}^N\left(-\theta^{(3)}_n -
\theta^{(4)}_n\right)=\gamma\,. \label{gamma}
\end{equation}
Here the variable $\theta^{(m)}_n$ is the phase difference across
the $m$-th junction in the $n$-th rhombus (see Fig. \ref{chain}).

This regular model was analyzed in details in ref. \cite{my}
within the limit $E_J\gg E_C$. At $E_C=0$ the phases
$\theta_n^{(m)}$ are classical variables which do not fluctuate.
Emerging classical states of the chain $\left|m,\{
\sigma_n^z\}\right>$ can be characterized by a set of binary
variables $\{\sigma_n^z\}$ (one for each rhombi) and an
integer-valued variable $m$. Note that each binary variable
$\sigma_n^z$ can be considered as $z$-projection of spin $\frac12$
ascribed to each rhombi. Energies of these classical states are
\begin{equation}
E_{m, \sigma}\approx \frac{E_J\sqrt2}{4N}(\widetilde{\gamma}-\pi
N/2-\pi S^z- 2\pi m)^2 -\sqrt{2}\delta S^z E_J+{\rm Const}.
\label{old E_under_flux_dif_from_pi}
\end{equation}
\begin{equation}
s_n^z=\frac{1}{2}\sigma_n^z\,, \qquad
S^z=\frac{1}{2}\sum_{n=1}^N\sigma_n^z\,, \qquad
\widetilde{\gamma}=\gamma+\frac{N\varphi}{2}\,.
\end{equation}

Finite $E_J/E_C$ gives rise to quantum phase slips (QPS) in the
chain, which mix different classical states leading to formation
of truly quantum ground state. Let us denote as $\upsilon$ the
amplitude of a QPS in one contact. At large $N \gg 1$ this
amplitude does not differ from the "spin flip"  amplitude for  a
single rhombus at $\Phi_r \approx \Phi_0/2$. In this approximation
we can use result from Ref.~\cite{IoffeFeigelman02}:
\begin{equation}
\upsilon\approx k
\left(E_J^3E_C\right)^{1/4}\exp\left(-1.61\sqrt{\frac{E_J}{E_C}}\right).
\label{upsilon_gen}
\end{equation}
Here $k$ is a numerical factor of order one which slightly depends on $E_J/E_C$ (see ref. \cite{my} for details).

In \cite{my} it was shown that calculation of the persistent
current in the large-$N$ limit can be reduced to the solution of a
Schr\"{o}dinger equation for a particle having a large spin
$S=\frac12 N$ moving in a periodic $\cos$-like potential, with
appropriate boundary condition.

More precisely, the quantum ground-state energy $E$ of the chain
in the limit $E_J\gg E_C$ can be obtained from solution of
Schr\"{o}dinger equation
\begin{equation}
\frac{\partial^2\psi}{\partial x^2} +(\widetilde{E}- 2 w \cos
2x\cdot
\sum_{n=1}^N\widehat{S}_n^x+2h\sum_{n=1}^{N}\widehat{S}_n^z)
\psi=0\,, \label{old_main_res_under_flux_dif_from_pi}
\end{equation}
where
\begin{equation}
\widetilde{E}=\frac{16NE}{\sqrt2 E_J \pi^2}\,, \qquad
w=\frac{64N\upsilon}{\sqrt2 E_J\pi^2}\,,\qquad
h=\frac{8N\delta}{\pi^2}\,. \label{bwh}
\end{equation}
Here $\upsilon$ --- amplitude for quantum phase slip in one rhombi
and $\delta=\pi-\varphi$. Function $\psi\equiv\psi(x,
\{\sigma_n\})$.  $\widehat{S}_n^x$ and $\widehat{S}_n^z$ are
standard operators of $x$ and $z$ projections of spin $\frac 12$
acting on $\sigma_n$. The magnetic flux inside the whole ring
enters the problem through the twisted boundary condition
\begin{equation}
e^{i\pi\widehat{S}^z+i\pi N/2}\psi\left(x+\pi/2,\sigma\right)
=e^{i\widetilde{\gamma}}\psi(x,\sigma)\,, \qquad
\label{old_boundary} \widetilde{\gamma}=\gamma+\frac{N\varphi}{2}
\end{equation}
In \cite{my} on the basis of this formulation of the problem
critical deflection $\delta\Phi^c$ was derived
\begin{equation}
\left(\delta\Phi^c\right)_{reg}\approx
0.2\left(\frac{\upsilon}{E_J}\right)^{2/3}\Phi_0\label{phi_c_reg}
\end{equation}

The aim of this section is to derive effective Hamiltonian similar
to (\ref{old_main_res_under_flux_dif_from_pi}) for the chain with
random fluxes $\Phi_r^n$ through rhombi or random stray charges.

To account for random stray charges we present the electrostatic
charging energy in the form
\begin{equation}
H_C=\sum_{n_1\,,\,n_2=1}^N\,\,\,\sum_{k_1\,,\,k_2=1}^3\frac12
\left[C^{-1}\right]^{k_1\,n_1}_{k_2\,n_2}
\left(Q_{n_1}^{(k_1)}-q_{n_1}^{(k_1)}\right)\left(Q_{n_2}^{(k_2)}-q_{n_2}^{(k_2)}\right)
\end{equation}
where $\left[C^{-1}\right]^{k_1\,n_1}_{k_2\,n_2}$ is the matrix of
inverse capacitances. Indices $n$ and $k$ numerates
superconducting islands ($3$ rows, $N$ islands in a row).
$Q_{n}^{(k)}$ --- charge of the $n$-th island in $k$-th row.
Parameters $q_{n}^{(k)}$ are determined by the random stray
charges.

Starting from this charging energy we may derive an additional
term in action (\ref{old_basic_action}) emerging from presence of
random stray charges (compare  with \cite{Larkin})
\begin{equation}
\delta S=-i\int
dt\sum_{n=1}^N\sum_{m=1}^4p_n^m\frac{d\theta_n^{(m)}}{dt}
\label{addition}
\end{equation}
Parameters  $p_n^m$ can be expressed in terms of charges
$q^{(k)}_n$. Corresponding expressions are a bit cumbersome and we
do not present them here. Below we write down some special
combinations of $p_n^m$ which we will need in our paper.

Additional term (\ref{addition}) has a form of total derivative.
Hence it does not change neither the classical states of the chain
nor the classical equations of motion and the real part of
classical action on a {\it single} tunneling trajectory. The only
effect of this term to give a tunneling amplitude along each path
its on phase factor. Since there are several QPS trajectories
between two classical states, all having the same real part of
tunneling action, these additional phase factors may give rise to
destructive interference of tunneling processes, leading to
reduction of total matrix element, connecting two classical
states.

Following ref. \cite{my} and taking into account complications due
to random stray charges we write tight-binding Hamiltonian

\begin{multline}
\widehat{H}|m, \sigma>= E_{m\sigma}|m,\sigma> + \frac{\upsilon}{2}
\sum_{n=1}^N\exp\left\{\frac{i\pi}{2}\left(3p_n^1-p_n^2-p_n^3-p_n^4
\right)\right\} \widehat{\sigma}_n^+|m,\sigma> +\\
\frac{\upsilon}{2}
\sum_{n=1}^N\exp\left\{\frac{i\pi}{2}\left(3p_n^2-p_n^1-p_n^3-p_n^4
\right)\right\} \widehat{\sigma}_n^+|m,\sigma> + \\
\frac{\upsilon}{2}
\sum_{n=1}^N\exp\left\{-\frac{i\pi}{2}\left(3p_n^3-p_n^1-p_n^2-p_n^4
\right)\right\} \widehat{\sigma}_n^+|m-1,\sigma> +\\
\frac{\upsilon}{2}
\sum_{n=1}^N\exp\left\{-\frac{i\pi}{2}\left(3p_n^4-p_n^1-p_n^2-p_n^3
\right)\right\} \widehat{\sigma}_n^+|m-1,\sigma> +h.\,c.
\label{H_acting_over_m_s}
\end{multline}
Performing Fourier transformation over variable $m$ according to
\begin{equation}
\left|x,\sigma\right>
=\sum_m\exp\left\{2i\left(2m-\frac{\widetilde{\gamma}}{\pi}+S^z+\frac{N}{2}\right)x\right\}
\left|m,\sigma\right>\,,\label{Fur'e}
\end{equation}
we obtain the effective Schr\"{o}dinger equation
\begin{equation}
\frac{\partial^2\psi}{\partial x^2} +\left(\widetilde{E}- 2 w \cos
2x \sum_{n=1}^{N}a_n\widehat{S}_n^x -  2 w \sin 2x
\sum_{n=1}^{N}b_n\widehat{S}_n^y +2h \widehat{S}^z\right)\psi=0\,,
\label{main_res_random_charge}
\end{equation}
and twisted boundary condition
\begin{equation}
e^{i\pi\widehat{S}^z+i\pi N/2}\psi\left(x+\pi/2,\sigma\right)
=e^{i\widetilde{\gamma}}\psi(x,\sigma)\,. \label{boundary}
\end{equation}
Here parameters $\widetilde{E}$, $w$ and $h$ are described by
equation (\ref{bwh}); $a_n$ and $b_n$ are random coefficients
which can be expressed in term of random charges $q_n^{(k)}$
\begin{equation}
a_n=\frac{\cos\pi Q_n^1+\cos\pi Q_n^2}{2}\,, \qquad
b_n=\frac{\cos\pi Q_n^1-\cos\pi Q_n^2}{2}
\end{equation}
\begin{equation}
Q_n^1\equiv
p_n^1-p_n^2=q_n^{(2)}-\frac{1}{3N}\sum_{n=1}^N\sum_{k=1}^3q_n^{(k)}
\qquad Q_n^2\equiv
p_n^3-p_n^4=q_n^{(3)}-\frac{1}{3N}\sum_{n=1}^N\sum_{k=1}^3q_n^{(k)}
\label{Q}
\end{equation}
Here we measure charges $q_n^{(k)}$ in units $2e$.

We turn now to generalization of (\ref{old_main_res_under_flux_dif_from_pi}) for the case of random
fluxes in rhombi. Imaginary-time action for this
problem is given by equation (\ref{old_basic_action}) but
additional conditions (\ref{varphi}) should be changed to
\begin{equation}
\sum_{m=1}^4\theta^{(m)}_n=\varphi_n\,, \qquad n=1,2,..,N\,,
\label{new_varphi}
\end{equation}
Following the same steps as before  we see that to
account for disorder in flux piercing   the rhombi one should just
replace the last term in
(\ref{old_main_res_under_flux_dif_from_pi}) by
\begin{equation}
2\sum_{n=1}^{N}h_n\widehat{S}_n^z
\end{equation}
Here $h_n=8N(\pi-\varphi_n)/\pi^2$.

Now we have effective Hamiltonians of the chain in presence of
disorder. In the rest of the paper we will  analyse these
Hamiltonians in order to find out influence of disorder on the
crossover point between $4e$- and $2e$-regimes.

\section{Influence of random charges on crossover point}
In this section we study influence of random stray charges on the
on pairing effect in rhombi chain. It is important to note that,
generally speaking,  even at  maximally frustrated point $h=0$ (in
contrast to regular chain) symmetry properties of Schr\"{o}dinger
equation (\ref{main_res_random_charge}) {\it do not} prohibit
$2e$-supercurrent. This is due to the fact that asymmetric
realizations of random charges with $q_n^{(2)}\neq q_n^{(3)}$ (and
thus $b_n\neq0$) break symmetry between two tunneling trajectories
of a Cooper pair within the same rhombus. Of course, random
charges preserve classical states of the rhombi chain and in a chain
with no QPS there would be no $2e$-supercurrent at maximally
frustrated point, but full quantum Hamiltonian of the chain does
not possess corresponding symmetry. Nevertheless, we will see
later that for typical realizations of random charges and at $\Phi_r=\Phi_0/2$
the $2e$-supercurrent  is small as compared to
$4e$-supercurrent. The reason for that is as follows: if at least
in one rhombi $b_n=0$ ( or $a_n=0$) then $2e$-supercurrent is
prohibited. Here we start with analysis of general situation of
rhombi chain with random charges and $h\neq0$ and then make some
conclusion on disordered rhombi chain in maximally frustrated
point.

We investigate the grand partition function for the system
described by equations (\ref{main_res_random_charge},
\ref{boundary})
\begin{equation}
Z=\int{\cal
D}x(\tau)\exp\left(-\int_0^\beta\frac{\dot{x}^2}{2}\right)\prod_{n=1}^N\Tr
\left[\widehat{U}_n(\beta)\right]\label{Z}
\end{equation}
with $\beta$ being the inverse temperature, $\beta\rightarrow
\infty$. Operators $\widehat{U}_n$ act in the space of spin
$\frac 12$ each  and are  functionals of $x(\tau)$ defined as
\begin{equation}
\frac{d \widehat{U}_n}{d\tau}=-\left(w f_n(\tau)
S^x+wg_n(\tau)\widehat{S}^y-h\widehat{S}^z\right)\widehat{U}_n
\label{defU}
\end{equation}
\begin{equation}
f_n(\tau)=a_n\cos 2x(\tau)\,, \qquad g_n(\tau)=b_n\sin 2x(\tau)
\end{equation}
As was shown in \cite{my}, in a regular chain with $E_J\gg E_C$
the borderline between $4e$- and $2e$-supercurrents is at rather
large $\Phi_r-\Phi_0/2$, in the sense that at the crossover point
$h\gg w$. So in this paper we will also consider this limit only.

Under such condition equation (\ref{defU}) can be  solved for
arbitrary functions $f(\tau)$ and $g(\tau)$. For
$\Tr\widehat{U}_n(\beta)$ we then find
\begin{multline}
\Tr \widehat{U_n}(\beta)=\exp\left(\frac{\beta
h}2+\frac{w^2}{4h^2}\left(f_n(0)-ig_n(0)\right)\left(f_n(\beta)+ig_n(\beta)\right)+\right.\\
\left. \frac{w^2}{8}\int_0^\beta d\tau_1 d\tau_2 \left(
f_n(\tau_1)f_n(\tau_2)+g_n(\tau_1)g_n(\tau_2)\right)e^{-h|\tau_1-\tau_2|}+
\right.
\\ \left.
i\frac{w^2}{4}\int_0^\beta d
\tau_1d\tau_2f_n(\tau_1)g_n(\tau_2)e^{-h|\tau_1-\tau_2|}\sign(\tau_1-\tau_2)\right)
\label{TRU}
\end{multline}
From equation (\ref{TRU}) we derive effective action for variable
$x$
\begin{equation}
Z=\int_0^\beta {\cal D}x(\tau) e^{-S[x(\tau)]}\qquad
S=S_{bound}+S_\tau
\end{equation}
\begin{equation}
S_{bound}=-\frac{w^2}{4h^2}\sum_{n=1}^N\left(a_n\cos
2x(\beta)+ib_n\sin 2x(\beta)\right) \left(a_n\cos
2x(0)-ib_n\sin2x(0)\right)
\end{equation}
\begin{multline}
S_\tau=\int_0^\beta d\tau\frac{\dot{x}^2(\tau)}{2}-
A\frac{w^2}{8}\int_0^\beta d\tau_1d\tau_2
\cos2x(\tau_1)\cos2x(\tau_2)\exp(|\tau_1-\tau_2|)-\\
B\frac{w^2}{8}\int_0^\beta
d\tau_1d\tau_2\sin2x(\tau_1)\sin2x(\tau_2)\exp(|\tau_1-\tau_2|)-
\\ iC\frac{w^2}{4h}\int_0^\beta d \tau_1d\tau_2\cos 2x(\tau_1)\sin2x(\tau_2)\exp(|\tau_1-\tau_2|)\sign(\tau_1-\tau_2)
\label{action_x}
\end{multline}
Where
\begin{equation}
A=\sum_{n=1}^N a_n^2\,, \qquad B=\sum_{n=1}^N b_n^2\,, \qquad
C=\sum_{n=1}^N a_n b_n \label{ABC}
\end{equation}

Note that similar approximation was used previously in \cite{my} for a regular chain.
In a regular chain semiclassical analysis which do not rely on linearization with respect to $w/h$ is also
possible. It turns out that exact value of the critical deflection $\delta\Phi^c$ differs only by $12\%$ from the one
obtained by linearization even for the case when at the crossover point $w/h=1$.

Action (\ref{action_x}) is nonlocal and looks a bit terrific, but
since
\begin{equation}
\left( \frac{\partial^2}{\partial
\tau_1^2}-h^2\right)e^{-h|\tau_1-\tau_2|}=-2h\delta(\tau_1-\tau_2)
\end{equation}
we can perform  Hubbard-Stratonovich  transformation and derive a
representation for grand partition function with local action
(after redefinition of time scale according to
$\tau\longrightarrow t/h$).
\begin{equation}
Z=\int_0^\beta {\cal D}x(\tau){\cal D}y(\tau){\cal D}z(\tau)
e^{-S[x(\tau)]}\qquad
\end{equation}
\begin{multline}
S=h\int_0^{\beta h}
d\tau\left(\frac{\dot{x}^2+\dot{y}^2+\dot{z}^2}2+\frac{y^2+z^2}{2}+
\alpha_1dy\cos2x+i\beta_1d\dot{y}\sin2x+\right.\\
\left. \alpha_2dz\sin2x+i\beta_2d\dot{z}\cos2x
-\frac{d^2}{2}(\beta_1^2\sin^2 2x+\beta_2^2 \cos^2 2x) \right)
\label{decoupling}
\end{multline}
Here $\alpha_1$, $\alpha_2$, $\beta_1$ and $\beta_2$ should
satisfy equations
\begin{equation}
\alpha_1^2+\beta_2^2=\frac{A}{N}\,, \qquad
\beta_1^2+\alpha_2^2=\frac BN\,, \qquad
\alpha_1\beta_1-\alpha_2\beta_2=\frac C N  \label{abab}
\end{equation}
and
\begin{equation}
d=\sqrt{\frac{w^2N}{2h^3}}=\frac{\sqrt{2}\pi}{\delta^{3/2}}\frac{\upsilon}{E_J}\label{d}
\end{equation}

As we see from equations (\ref{abab}) we have some freedom in
definitions of these parameters. Namely, if we set
\begin{equation}
\alpha_1=\sqrt{\frac{A}{N}}\sin\kappa_1\,, \qquad
\beta_2=\sqrt{\frac{A}{N}}\cos\kappa_1
\end{equation}
\begin{equation}
\alpha_2=\sqrt{\frac{B}{N}}\sin\kappa_2\,, \qquad
\beta_1=\sqrt{\frac{B}{N}}\cos\kappa_2
\end{equation}
then (\ref{abab}) reduces to
\begin{equation}
\sin(\kappa_1-\kappa_2)=\frac{C}{\sqrt{AB}}
\end{equation}
and we have only one equation for two parameters $\kappa_1$ and
$\kappa_2$. This freedom will not be important for us. In present
paper for the reasons described below we will mostly consider
action  (\ref{decoupling}) under condition $C=0$ and chose
\begin{equation}
\alpha_1=\sqrt{\frac{A}{N}}\,, \qquad
\alpha_1=\sqrt{\frac{A}{N}}\,, \qquad \beta_1=0\,, \beta_2=0
\end{equation}

 The main feature of the result presented above is that we
have reduced original problem including large number of random
variables ($\sim N$) to a problem with only {\it three } random
parameters $A$, $B$, and $C$. More over, since $A$, $B$ and $C$
arise as sums of large number $N$ of independent random variables
(see equation (\ref{ABC})), it is natural to expect that they obey
Gaussian statistics. Let us assume that parameters $Q_n^1$ and
$Q_n^2$ in (\ref{Q}) are uniformly distributed on $[-1, 1]$. It
corresponds to strong fluctuations of stray charges from sample to
sample. Under such an assumption one can easily find
\begin{eqnarray}
\left<A\right>=\left<B\right>=\frac{N}{4}\nonumber\\
\left<A^2\right>=\left<B^2\right>=\frac{N^2}{16}+\frac{5N}{64}\,,
\qquad \left<AB\right>=\frac{N^2}{16}-\frac{3N}{64}
\label{stat_ABC}\\
\left<C\right>=0\,, \qquad \left<C^2\right>=\frac{N}{64}\nonumber
\end{eqnarray}

Our strategy in the rest part of this section will be to analyse
properties of the system described by equations (\ref{decoupling},
\ref{boundary}) with {\it fixed } $A$, $B$ and $C$ and then make
some statistical analysis. Under fixed $A$, $B$ and $C$ the
main subject of our investigation will be whether the system is in
regime of dominating $4e$-supercurrent or not.

First of all let us  analyze action (\ref{decoupling}) for a
trajectory where $x(\tau)$, $y(\tau)$ and $z(\tau$) are constant.
We have
\begin{equation}
S_{st}=\beta h\left(\frac{y^2+z^2}{2}+ \alpha_1dy\cos2x+
\alpha_2dz\sin2x-\frac{d^2}{2}(\beta_1^2\sin^2 2x+\beta_2^2 \cos^2
2x) \right)
\end{equation}

This static action has two groups of minima (we call them even and
odd, suppose $A>B$)
\begin{equation}
x=\pi n\,, \qquad y=-\alpha_1 d \,, \qquad z=0
\end{equation}
and
\begin{equation}
x=\frac{\pi}2+\pi n\,, \qquad y=\alpha_1 d \,, \qquad z=0
\end{equation}
where $n$ is an arbitrary integer. All these minima correspond to
the same value of $S_{st}$. So we have to consider two types of
tunnelling trajectories. Trajectories of the first type connect
minima of the same group, i.e. "even-even" and "odd-odd", and
corresponding variation of the variable $x$ between minima is $\pm
\pi$, whereas $y$ returns to its original value. Trajectories of
the second type connect minima of opposite parity (i.e. opposite
signs of $y$), and change $x$ variable by $\pm \frac{\pi}2 $. It
is not difficult to see from
Eqs.(\ref{Fur'e},\ref{main_res_random_charge},\ref{boundary} ),
 that increment $\Delta x$ of the variable $x$ along tunnelling trajectory is in one-to-one
correspondence to the elementary charge transported along the
rhombi chain: $q_0 = \frac{4e}{\pi}\Delta x$. Therefore
trajectories of the first type lead to $4e$ - supercurrent,
 whereas trajectories of the second type produces usual $2e$-quantized
supercurrent. In semiclassical approximation  amplitudes of the
supercurrent components are determined primarily by the classical
actions on corresponding trajectories:
\begin{equation}
I(\gamma) = I_{2e}\sin\widetilde{\gamma} +
I_{4e}\sin(2\widetilde{\gamma}), \label{gen_cur}
\end{equation} where
\begin{equation}
I_{4e} = A_{4e}  \exp(-S_E^{4e}),  \quad  I_{2e} = A_{2e}
\exp(-S_E^{2e}) \, , \label{components}
\end{equation}

To find out whether $4e$-supercurrent dominates in the system, we
should compare classical actions on the trajectories of two types
described above. So we examine classical equations of motion for
action (\ref{decoupling}). For simplicity here we put coefficient
$C$ to zero. It can be shown that relatively small $C$ of order
$\sqrt{N}$ (compare to (\ref{stat_ABC})) is not important for our
future purpose.
\begin{equation}
\ddot{x}+2\alpha_1 d y \sin2x-2\alpha_2 d z \cos 2x =0
\label{equation_x}
\end{equation}
\begin{equation}
\ddot{y}-y=\alpha_1 d \cos 2x \label{equation_y}
\end{equation}
\begin{equation}
\ddot{z}-z=\alpha_2 d \sin 2x \label{equation_z}
\end{equation}

Here we will analyse equations (\ref{equation_x},
\ref{equation_y}, \ref{equation_z}) and find analytic expression
for the borderline between $2e$- and $4e$-regimes under conditions
\begin{equation}
Ad^2/N\gg1\,,\qquad   A-B\ll A\label{condAB}
\end{equation}
We also present results of numerical computation of the borderline
in general situation.

Let us start with $2e$-trajectory. Note that for variables $x$ and
$y$ characteristic frequency is 1. Let us suppose that on
$2e$-trajectory $x$ varies slowly so that characteristic frequency
$\omega_x\ll 1$. Than we can eliminate $y$ and $z$ from equations
of motion in adiabatic approximation and obtain
\begin{equation}
\ddot{x}-d_1^2\sin4x=0 \label{slow_x}
\end{equation}
\begin{equation}
d_1^2=\frac{(A-B)d^2}{N+2d^2(A+B)}
\end{equation}
We see that under conditions  (\ref{condAB}) $\omega_x\sim d_1$ is
indeed small. Equation (\ref{slow_x}) has solution corresponding
to $2e$-trajectory.
\begin{equation}
x(\tau)=\frac{1}{2}\arccos{\left(-\tanh\left(2d_1\tau\right)\right)}\,,
\label{slow_x_sol}
\end{equation}
Classical action on this solution is
\begin{equation}
S_{2e}=h\sqrt{\frac{d^2(A-B)}{N}\left(1+2\frac{(A+B)d^2}{N}\right)}
\label{two_e_action}
\end{equation}

Let us now turn to examination of $4e$-trajectories. Here we
suppose that $x$ is a  fast variable i.e. $\omega_x\gg 1$. We can
neglect then variation of $y$ and $z$ on classical trajectory and
put $y=-\alpha_1d$, $z=0$. This is consistent with boundary
conditions for $4e$-trajectory. Equation for $x$ becomes
\begin{equation}
\ddot{x}-\frac{2A}{N}\, d^2 \sin2x =0 \label{fast_x}
\end{equation}
Again, due to (\ref{condAB}),  we see that $x$ indeed varies fast
on $4e$-trajectory since $\omega_x\sim \sqrt{Ad^2/N}\gg1$. So we
find $4e$-trajectory and corresponding action
\begin{equation}
x(\tau)=\arccos{\left(-\tanh\left(2\sqrt{\frac{Ad^2}{N}}\tau\right)\right)}\,,
\label{x_0_reg2}
\end{equation}
\begin{equation}
S_{4e}=4h\sqrt{\frac{Ad^2}{N}} \label{four_e_action}
\end{equation}

Comparing equations (\ref{two_e_action}) and (\ref{four_e_action})
we  find the line of crossover between $4e$- and $2e$- regimes:
the set of points $(A, B)$ such that $S_{4e}=S_{2e}$.
\begin{equation}
\frac{Bd^2}{N}=\sqrt{\left(\frac{Ad^2}{N}\right)^2-8\frac{Ad^2}{N}}\label{cross_line}
\end{equation}

The result presented above was derived under assumption $C=0$ but
proceeding in the same way with $C\lesssim \sqrt{N}$ one can show
that at large $N$ nonzero $C$ does not affect the crossover line
(\ref{cross_line}). Therefore in the rest part of this paper we
will completely ignore coefficient $C$.

\begin{figure}[t] \centering
\psfrag{comb1}[cr][c][1.15][-90]{\tiny $\frac{\delta\Phi}{(\delta\Phi^c)_{reg}}\widetilde{B}$}
\psfrag{comb2}[tl][c][1.15]{\tiny $\frac{\delta\Phi}{(\delta\Phi^c)_{reg}}\widetilde{A}$}
\psfrag{subplotb}[bc][c]{\small b)}
\psfrag{u}[c][c]{\small $u$}
\psfrag{v}[c][c]{\small $v$}
\psfrag{a}[bc][c]{\small a)}\hspace{-1cm}
\includegraphics[width=200pt]{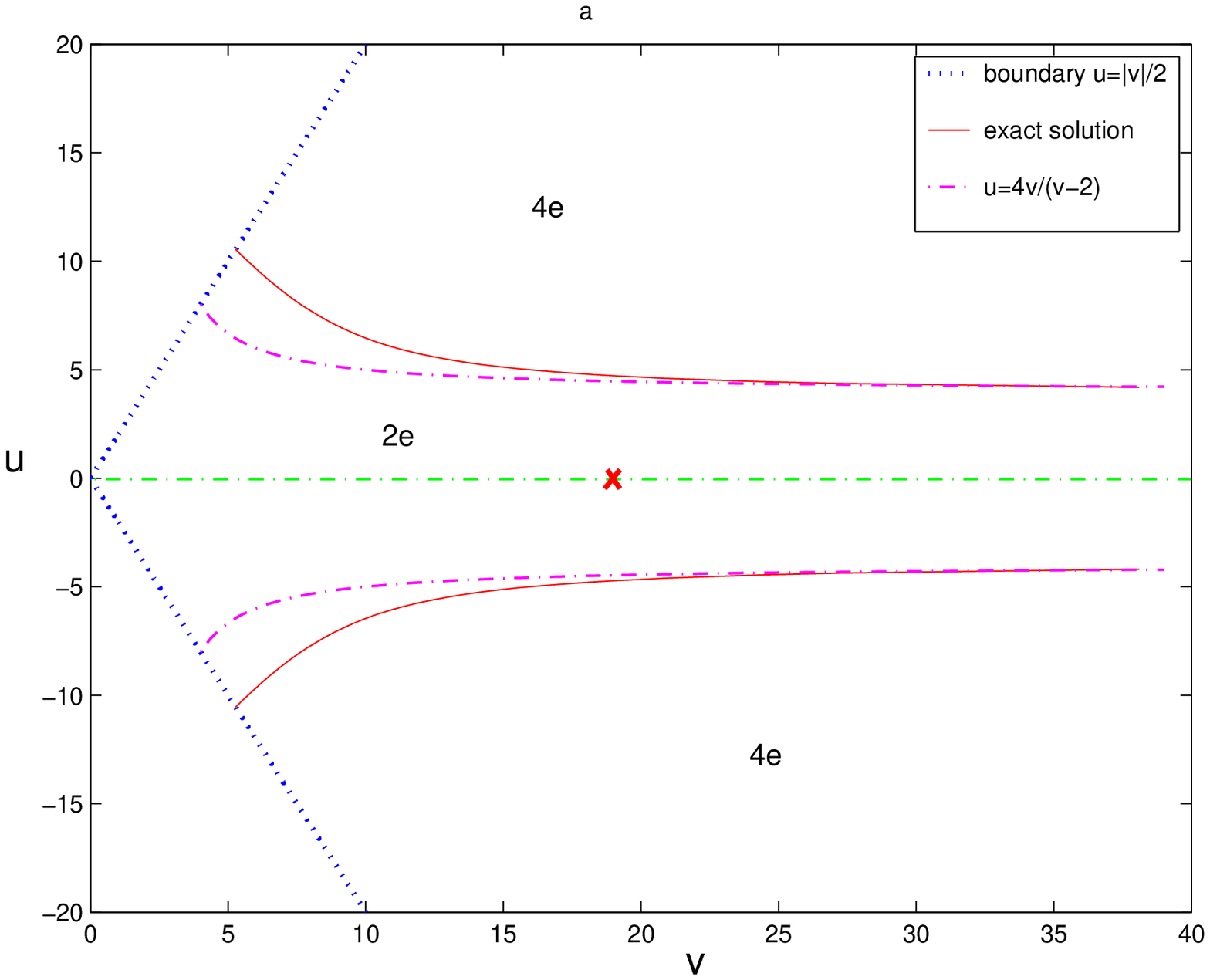}\hspace{2cm}
\includegraphics[width=170pt]{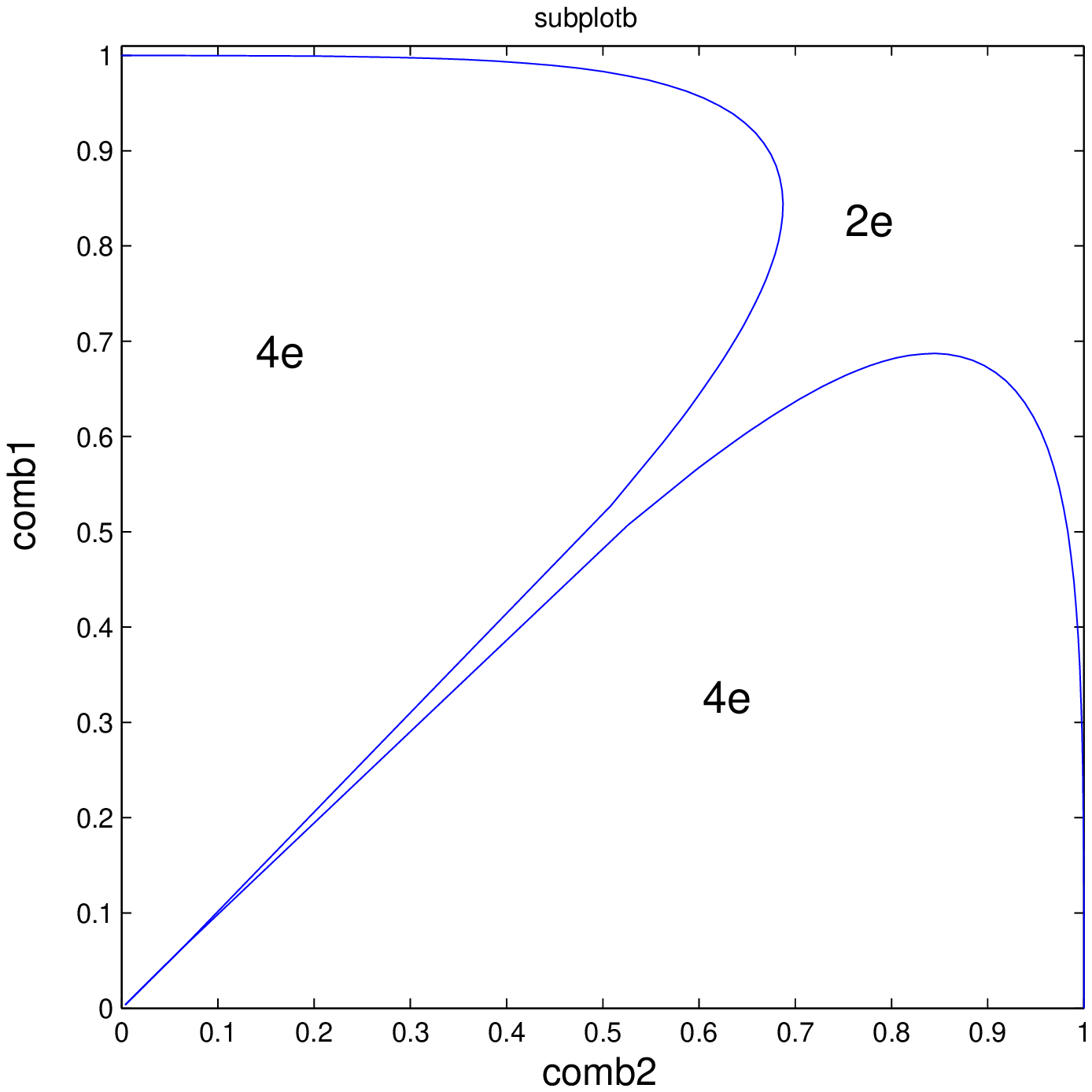}
 \caption{\small "Phase diagram" of disordered chain. Solid lines on the both figures mark the crossover
$2e$- and $4e$-regimes. Note that these lines does not correspond to any phase transition. The crossover however is sharp
at large $N$ since actions $S_{4e}$ and $S_{2e}$ are proportional to the number of rhombi. Subplot a) presents crossover
line in variables $u$ and $v$ defined by (\ref{def_uv}) which are useful for calculations. Subplot b) depicts the same
in  a bit more intuitive way. Here we introduce factors
$\widetilde{A}=\left(\frac{NA}{A^2+B^2}\right)^{1/3}$ and  $\widetilde{B}=\left(\frac{NB}{A^2+B^2}\right)^{1/3}$
describing realization of disorder; $\left(\delta\Phi^c\right)_{reg}$ --- critical deviation for a clean chain with same
$E_J/E_C$.}
\label{crossover_fig}
\end{figure}

The borderline obtained from numerical solution of classical
equations of motion (\ref{equation_x}, \ref{equation_y},
\ref{equation_z}) and its asymptotic form (\ref{cross_line}) is
presented on figure \ref{crossover_fig}.
Instead of $A$ and $B$ we have used here more convenient variables
\begin{equation}
u=d^2\frac{(A-B)}{N}\,, \qquad v=\frac{d^2(A+B)}{2N}
\label{def_uv}
\end{equation}
Note that by definition $v>|u|/2$. From
(\ref{cross_line}) we see that in terms of $u$ and $v$ at large $v$ the borderline between $2e$- and
$4e$-regimes is given by
\begin{equation}
u=f(v)\approx\frac{4v}{v-2}\rightarrow4\,, \qquad v\rightarrow \infty
\end{equation}
On figure \ref{crossover_fig}b $\,$ "phase diagram" of the disordered chain is presented in a more intuitive manner:
we emphasize here that $4e$-supercurrent exist in a small vicinity of the maximally frustrated point.

Equipped with this result,  for any {\it given} set of quenched
random charges (characterized by coefficients $A$ and $B$) we can
(in principle) determine whether $4e$-supercurrent dominates in
the chain. Of course experimentally we have no access to such
quantities as $A$ and $B$. Hence we need some statistical
description of rhombi chain with random charges. Such a
description is provided by the probability ${\cal P}(E_J/E_C, N,
\delta)$ to find dominating $4e$-supercurrent in the system.

Assuming Gaussian statistics for $u$ and $v$  and taking into
account (\ref{stat_ABC}) one can derive probability distribution
for $u$ and $v$
\begin{equation}
P(u, v)=\frac{8 N}{\pi d^4} \exp\left(-\frac{2 N
u^2}{d^4}\right)\exp\left(-\frac{32N(v-v_0)^2}{d^4}\right)
\label{puv}
\end{equation}
\begin{equation}
v_0=\frac{d^2}{4}
\end{equation}

Required  probability can be written as
\begin{equation}
{\cal P}_{4e}(E_J/E_C, N, \delta)=1-2\int_0^{+\infty} dv
\int_0^{f(v)} du  P(u, v)\label{prob1}
\end{equation}

Maximum of probability distribution (\ref{puv}) lies at $u=0$,
$v=d^2/4$. This point for $d=9$ is marked on figure
\ref{crossover_fig} with a cross. At sufficiently large $d$ we can
replace $f(v)$ in (\ref{prob1}) by $4$. We then get
\begin{equation}
{\cal P}_{4e}(E_J/E_C, N,
\delta)=1-\frac{2}{\sqrt{\pi}}\int_0^{\sqrt{\frac{32N}{d^4}}} du\,
\exp\left(-u^2\right)=1-\Erf\left(\sqrt{\frac{32N}{d^4}}\right)
\label{P4e_approx}
\end{equation}

Let us introduce parameter $\kappa$ and {\it define} critical
deviation from the maximally frustrated point $\delta \Phi_c$ as
deviation under which the probability (\ref{P4e_approx}) equals
$\kappa$. Reasonable choice for $\kappa$ is $0.5$ or $0.75$ or
something else. From (\ref{P4e_approx}) we then get central result
of this section
\begin{figure}
\psfrag{P}[c][c]{${\cal P}_{4e}$}
\psfrag{delta}[tc][c]{$\frac{\delta\Phi^c}{\Phi_0}\times 10^3$}
\psfrag{mya}[c][c]{a)}
\psfrag{myb}[c][c]{b)}
\psfrag{myc}[c][c]{c)}
\psfrag{myd}[c][c]{d)}

\hspace{-1cm}\includegraphics[width=200pt]{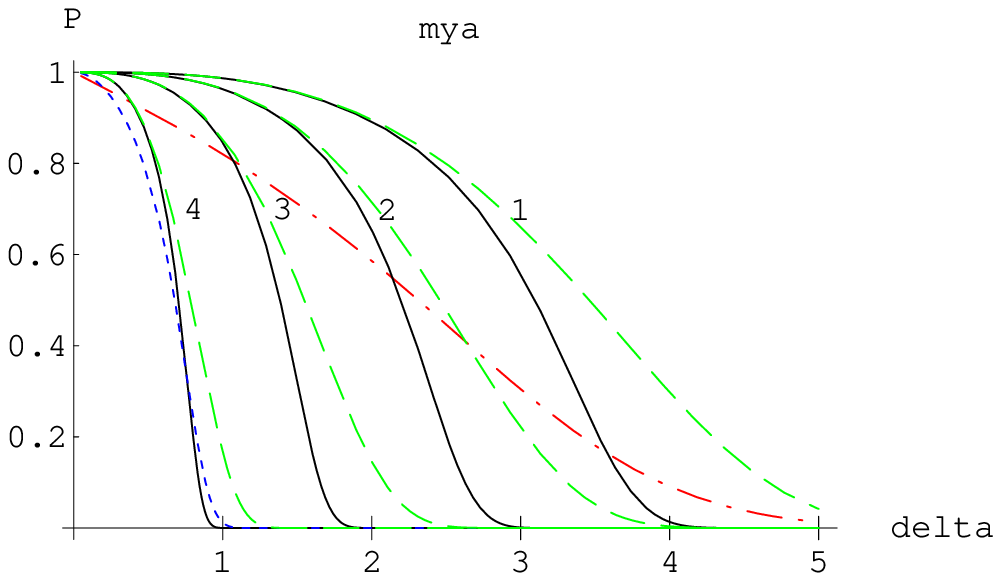}\hspace{2cm}
\includegraphics[width=200pt]{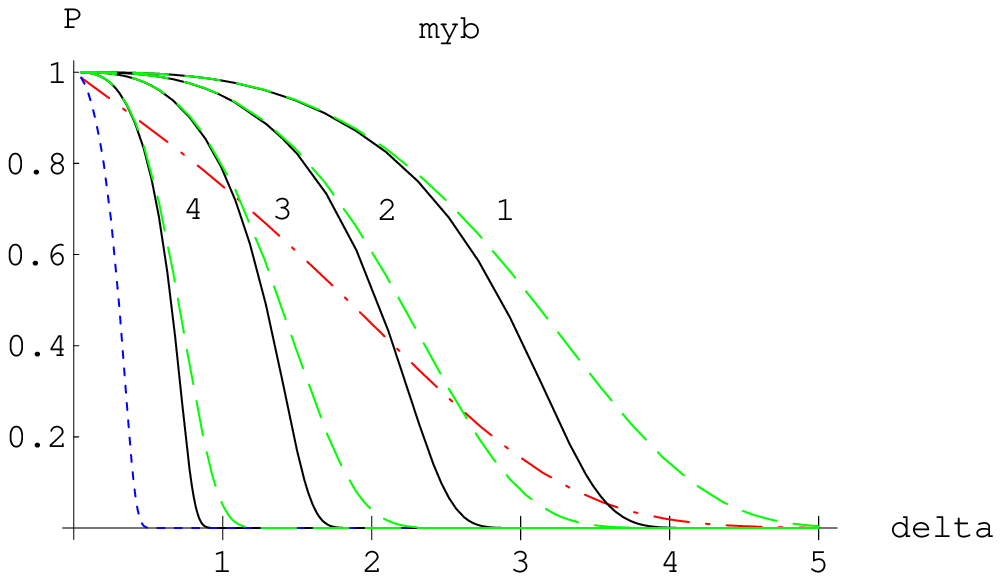}

\hspace{-1cm}\includegraphics[width=200pt]{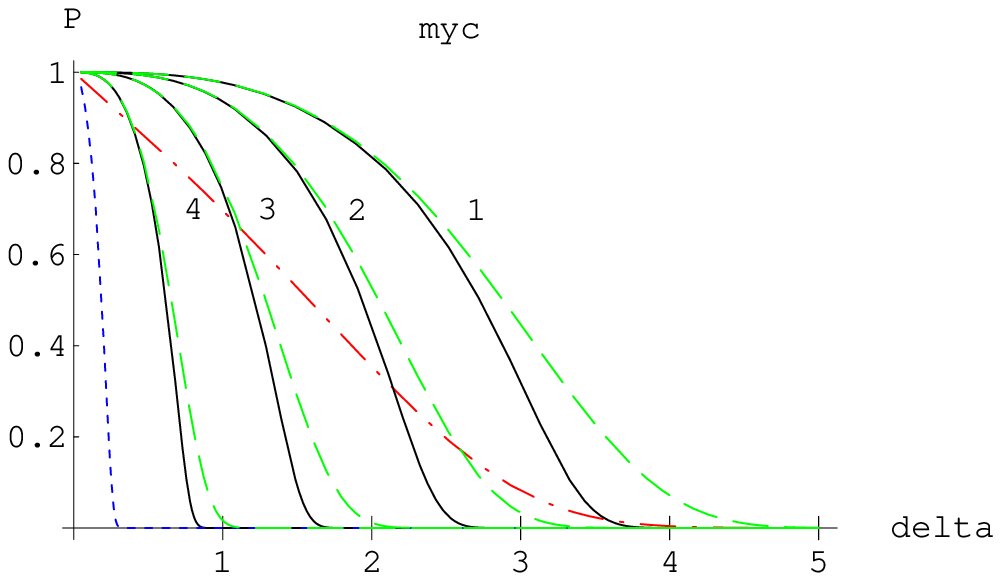}\hspace{2cm}
\includegraphics[width=200pt]{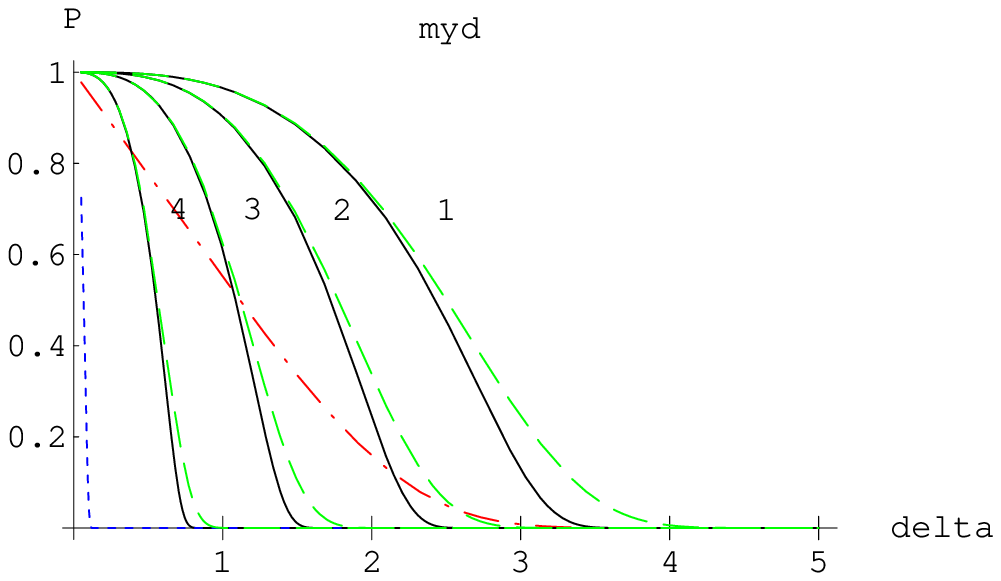}
 \caption{\small Probability ${\cal P}_{4e}$ as function of deviation from maximally frustrated point.
Subplots a), b), c), d),
correspond to $N=10,\, 20,\, 30,\, 70$.}
\label{p4e}
\end{figure}

\begin{equation}
\frac{\delta
\Phi_c}{\Phi_0}=\frac{\left(\Erf^{-1}\left(1-\kappa\right)\right)^{1/6}}{2^{3/2}\pi^{1/3}}\frac{1}{N^{1/6}}
\left(\frac{\upsilon}{E_J}\right)^{2/3}\label{phi_c_random_charges}
\end{equation}
Comparing (\ref{phi_c_random_charges}) and (\ref{phi_c_reg}) we
see that for any experimentally reasonable $N$ critical deviation
from the maximally frustrated point does not differ from the one
in a regular chain.
However one should remember that theory
presented above has several limitations. To clarify this
question we present here several graphics for ${\cal
P}_{4e}(\delta\Phi, E_J/E_C, N)$ at different $E_J/E_C$ and $N$
together with the boundaries of the validity region of the
approximation used (fig. \ref{p4e} ). Subplots a), b), c), d)
correspond to different number of rhombi $N=10, 20, 30, 70$
respectively. Solid and dashed lines $1, 2, 3, 4$ correspond to
different $E_j/E_C=10, 12, 15, 20$. Dashed lines were obtained
with the aid of equation (\ref{P4e_approx}) relying on the
condition $d^2>1$ whereas solid lines were produced by exploiting
equation (\ref{prob1}) with $f(v)$ determined from numerical
calculations.  The validity region should be determined as
follows. First of all, we have used assumption $h>>w$. Taking
$h=w$ as a criterium for the edge of our validity region one finds
the dash-doted curves on figure \ref{p4e}. Strictly speaking only
those parts of solids curves are valid, which lie under the
corresponding dash-doted curves. Now we have to ensure that rhombi
chain is in quantum regime, i.e. currents are exponentially small.
Simple estimates show that this is true above the doted line on
figure \ref{p4e}. At large $N$ this limitation is very soft.
Finally, our approximations are not valid at probabilities
${\cal P}_{4e}$ very small or very close to unity since they rely on
gaussian statistics for quantities $A$ and $B$ which certainly
does not describe rare events.

Qualitatively, results presented above can be interpreted as
follows: at large $E_J/E_C$ phase fluctuations in a single rhombi
are weak and this makes random charges inefficient, however in a
long chain global supercurrent is still exponentially suppressed.
From the analysis presented above one concludes that we can
guarantee with high probability existence of dominating,
suppressed by quantum fluctuations,  $4e$-supercurrent in long
chain ($N\sim 20$) with large $E_J/E_C\sim 20$ and for such a
chain critical deflection $\delta\Phi^c/\Phi_0\sim 10^{-3}$. If we
compare this result with critical deflection
$(\delta\Phi^c)_{reg}/\Phi_0\approx 1.2 \times 10^{-3}$ for a
regular chain with the same $E_J/E_C$ we conclude that the
influence of disorder on pairing effect is really small. On the
other hand in a regular chain it was possible to  obtain critical
deflection of order $10^{-2}$ by choosing $E_J/E_C\sim 8$.
For this set of parameters we cannot,  use our theory
quantitatively; we expect however that qualitatively the same
behaviour as described above for larger values $N$ and $E_J/E_C$ should be observed
here as well.

\section{Modulation of the supercurrent by capacitive gate}
In this section we will discuss briefly influence of regular gates on pairing effect
in maximally frustrated rhombi chain. We suppose that we have a gate, which is capacitively connected to all
superconducting islands in one row (fig. \ref{chain}).

Let us consider first rhombi chain with no disorder. Our system still can be described by equations
(\ref{main_res_random_charge}, \ref{boundary}), but now $a_n$ and $b_n$ are no more random. This coefficients
can be expressed in terms of the gate voltage according to
\begin{equation}
a_n=\frac{1+\cos\pi C_gV_g}{2}\equiv a\,, \qquad
b_n=\frac{1-\cos\pi C_gV_g}{2}\equiv b
\end{equation}
We will show here that the supercurrent in the chain is rather sensitive to the gate voltage.
Since the gate voltage enters the problem in a particularly quantum way (via phases of virtual QPS), this
dependence may provide an experimental test for the quantum nature of chain's state near the maximally frustrated point.

The problem we are considering now is much simpler as compared to the problem of random stray charges discussed above.
Now our Hamiltonian commutes with the total spin of the chain and we can apply semiclassical approximation directly
(e.g. by introducing spin coherent state path integral,  c.f. \cite{my, Klauder}). It's easy to see that the ground state
corresponds to the maximal total spin $S=N/2$. After proper redefinition of the time scale we can write imaginary-time
action in the form
\begin{equation}
S_E=\frac{N}{2}\int d\tau\left(-i\cos\theta\dot{\phi}
+\frac{w\dot{x}^2}{N}+\left(a\cos2x\cos\phi+b\sin2x\sin\phi\right)\sin\theta\right)=N\widetilde{S}_E(E_J/E_C,\, C_gV_g)
\label{S_reg_gate}
\end{equation}
Here angles $\theta$ and $\phi$ parameterize coherent states of the spin.
Again one easily find two types of tunneling trajectories corresponding to $4e$- and $2e$-supercurrents respectively
(e.g. $4e$-trajectory which connects $\left|x=0,\, \theta=\pi/2,\, \phi=-\pi\right>$ with
$\left|x=\pi,\, \theta=\pi/2,\, \phi=-\pi\right>$ and $2e$-trajectory connecting
$\left|x=0,\, \theta=\pi/2,\, \phi=-\pi\right>$
with $\left|x=\pi/2,\, \theta=\pi/2,\, \phi=0\right>$).
Action $\widetilde{S}_E$ is a function of $E_J/E_C$ and $C_g V_g$ and does not depend on the number of rhombi.

From (\ref{S_reg_gate}) we have classical equations of motion (it is convenient to change variables according to
$\theta \rightarrow \pi/2 +i\theta$)
\begin{eqnarray}
\frac{w}{N}\ddot{x}+(a \sin2x\cos\phi-b\cos2x\sin\phi)\cosh\theta=0\\
\dot{\theta}-(a\cos2x\sin\phi-b\sin2x\cos\phi)=0\\
\cosh\theta\, \dot{\phi}-(a\cos2x\cos\phi+b\sin2x\sin\phi)\sinh\theta=0
\end{eqnarray}

Note that $w/N\sim\upsilon$ is a very good small parameter at $E_J\gg E_C$. It allow us to obtain actions
 $\widetilde{S}_{4e}$ and
$\widetilde{S}_{2e}$ analytically. Characteristic frequency of variable $x\,$ is $\omega_x\sim \sqrt{N/w}\gg1\,$,
while for spins such a frequency is $\omega_s\sim 1$. This means that on
$4e$-trajectory, which does not require any change in spin, we can assume spin to be constant (at least while
coefficients $a$ and $b$ are not too close).
This immediately leads to
\begin{equation}
S_{4e}\approx \frac{16 N}{2^{1/4}\pi}\sqrt{\frac{\upsilon}{E_J}\left(1+\cos\pi C_g V_g\right)}
\label{S_4e_gate}
\end{equation}
On the other hand, on $2e$-trajectory smallness of $w/N$  allows us to omit $w\ddot{x}/N$ in equation for $x$.
Resulting equations can be integrated analytically and we obtain $2e$-action which, practically does not depend on
$E_J/E_C$
\begin{equation}
S_{2e}\approx N\arcsh\frac{\sqrt{\left|a^2-b^2\right|}}{b}=
N\arcsh\frac{2\sqrt{\left|\cos{\pi C_g V_g}\right|}}{1-\cos \pi C_g V_g}
\label{S_2e_gate}
\end{equation}

\begin{figure}
\psfrag{S4e}[c][c]{\tiny $\widetilde{S}_{4e}$}
\psfrag{S2e}[c][c]{\tiny $\widetilde{S}_{2e}$}
\psfrag{CV}[tc][c]{\tiny $C_g V_g/2e$}
\psfrag{ma}[bc][c]{\small a)}
\psfrag{myb}[bc][c]{\small b)}
\includegraphics[width=210pt]{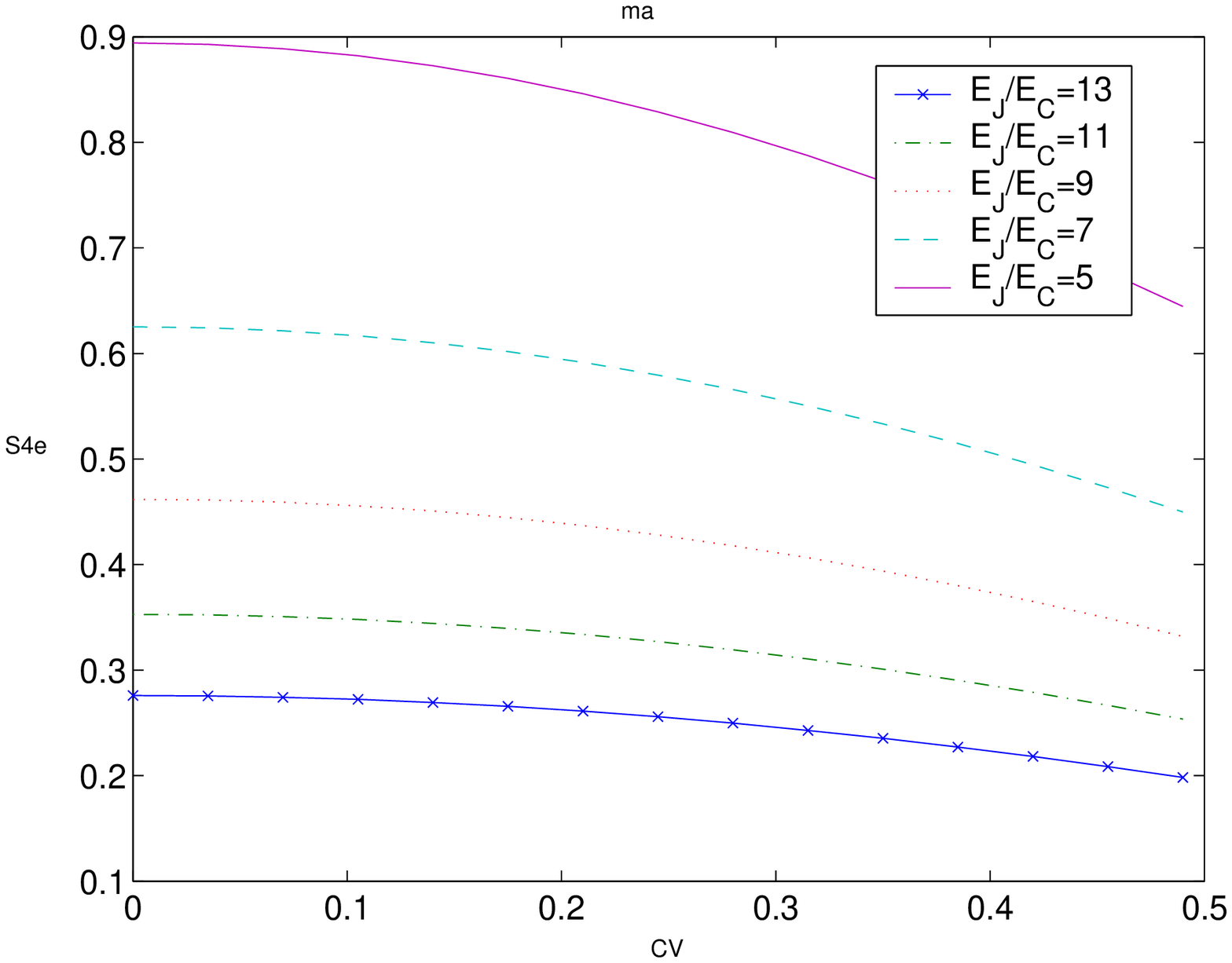}\hspace{1cm}
\includegraphics[width=210pt]{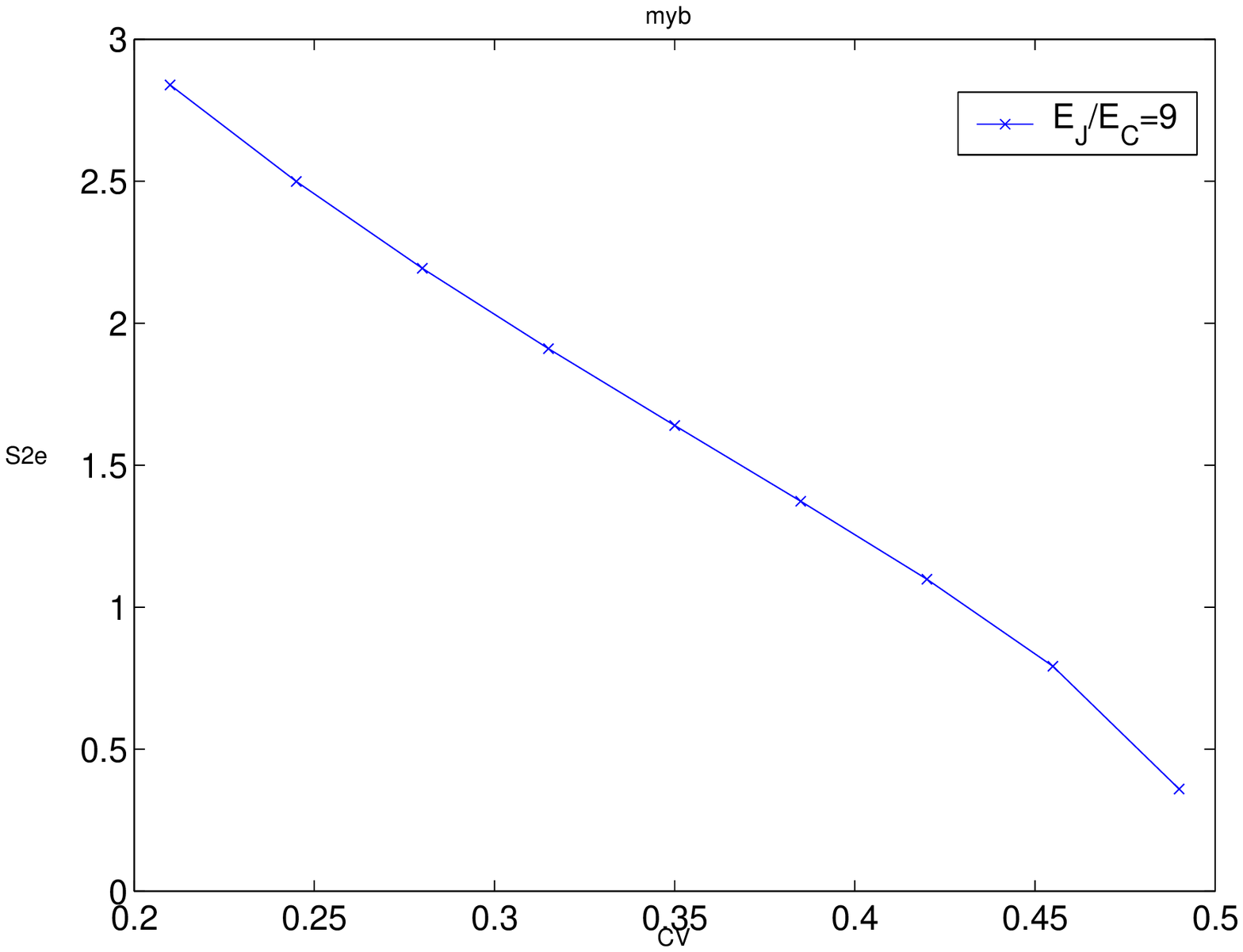}
 \caption{\small Modulation of $4e$- and $2e$-supercurrent in a clean chain by external gate. }
\label{S4e2e_gate_fig}
\end{figure}

Actions $\widetilde{S}_{4e}$ and $\widetilde{S}_2e$ can also be evaluated numerically.  Results of numerical
calculations are presented on figure \ref{S4e2e_gate_fig}.  Analytical results
(\ref{S_4e_gate}, \ref{S_2e_gate}) are in very good agrement with numerical data. Note that $2e$-supercurrent
dominates over the current of pairs of Cooper pairs only in small vicinity of the point
$C_g V_g/e=1$.

Note that in a regular chain change in $4e$-action upon applying external gate voltage is negative and proportional
to the number of rhombi, i.e. external gate leads to significant increase of otherwise  suppressed by fluctuations
$4e$-supercurrent.

Let us now discuss influence of a regular gate on a chain with random stray charges.
We will treat this problem within the same limit as in the previous section: we consider chain which is rather
far from the maximally frustrated point in the sense  $h\gg w$. From the preceding  analysis we know that $4e$-action
is given by equation (\ref{four_e_action}). Coefficient $A$ acquires now the following form
\begin{equation}
A=\frac{1}{4}\sum_{n=1}^{N}\left(cos\pi Q_n^1+\cos\pi(Q_n^2+C_g V_g)\right)^2=A_0+\Delta A
\end{equation}
Here $A_0$ is the value of $A$ at zero external bias.
Assuming that fluctuations of random charges are strong one easily finds
\begin{equation}
\left<\Delta A\right> =0\, \qquad \left<\Delta A ^2\right>=\frac{N}{32}\left(\sin^2\pi C_g V_g+
8\sin^2\frac{\pi C_g V_g}{2}\right)
\end{equation}
So typical change in $4e$-action can be estimated as
\begin{equation}
\delta S_{4e}\sim \frac{4\sqrt{2}\,d\,\delta}{\pi^2}\,\sqrt{N\left(\sin^2\pi C_g V_g+
8\sin^2\frac{\pi C_g V_g}{2}\right)}
\end{equation}
This result differs from the analogous one for a regular (without offset charges)
chain in two important aspects: i) we see that
change in action is now proportional to $\sqrt{N}$ instead of $N$; ii) variation of the
tunnelling action can now be both negative or positive, i.e. applying external voltage we
can {\it decrease} $4e$-supercurrent as well as increase it, depending
on the realization of charge disorder.
From the above analysis we conclude that in a disordered chain current modulation
by external gates is random and much weaker than in a regular chain,
but it is still present and can be used to demonstrate quantum coherence of the
tunnelling processes which occur in the chain.

\section{Influence of "magnetic disorder" upon the crossover point}
We now consider the effect of  randomness in the values of magnetix fluxes through
different rombi. Just as in the previous
section we start from the grand partition function given by
(\ref{Z}) with
\begin{equation}
\frac{d \widehat{U}_n}{d\tau}=-\left(w\, \cos2x(\tau)
S^x-h_n\widehat{S}^z\right)\widehat{U}_n \label{defU_random_h}
\end{equation}

We presume that fluxes $\Phi_r^n$ are uniformly distributed on
$(\Phi_r-\Delta\Phi, \Phi_r+\Delta\Phi)$, i.e. probability
distribution for $h_n$ is
\begin{equation}
P(h)=\left\{
\begin{array}{c}
\frac{1}{2\sigma}\,, \qquad  |h-h_0|<\sigma\\
0\,, \qquad |h-h_0|>\sigma
\end{array}\right.
\end{equation}
where
\begin{equation}
h_0=\frac{16  N (\Phi_r-\Phi_0/2)}{\pi\Phi_0}\,, \qquad \sigma=
\frac{16  N \Delta\Phi}{\pi\Phi_0}\,.
\end{equation}
Actually particular form of $P(h)$ is not important for us as we
assume that $\sigma\ll h_0$ and use perturbation theory in
$\sigma/h_0$. We are interested in the critical deviation from the
maximally frustrated point $\Phi_r=\Phi_0/2$ destroying
$4e$-supercurrent. Therefore we presume that $h_0\gg w$. In such
an approximation we can use equation (\ref{TRU}) to calculate $\Tr
\widehat{U}_n(\beta)$ and get effective action for variable $x$
\begin{equation}
S=\int_0^\beta d\tau\frac{\dot{x}^2(\tau)}{2}- \frac{w^2 h
N}{4}\int \cos2x(\tau_1)D(\tau_1-\tau_2)\cos2x(\tau_2)
\label{random_h_action}
\end{equation}
\begin{equation}
D(\tau_1-\tau_2)=\frac{1}{2 h_0 N}\sum_{n=1}^N
\exp\left(-h_n|\tau_1-\tau_2|\right)\label{D}
\end{equation}

At large $N$ we can replace $D(\tau_1-\tau_2)$ in
(\ref{random_h_action}) by its mean value
\begin{equation}
\overline{D}(\tau_1-\tau_2)=\frac{1}{2h_0}\int dh\,
e^{-h|\tau_1-\tau_2|}P(h)
\end{equation}
Fourier transformation of $\overline{D}$ for small $\sigma/h_0$
reads
\begin{equation}
\overline{D}(\omega)=\frac{1}{h_0^2+\omega^2}\left(1+ \sigma^2
\frac{h_0^2-3\omega^2}{3(h_0^2+\omega^2)^2}\right)
\end{equation}
Again, using the fact that
\begin{equation}
\overline{D}^{-1}(\omega)=\omega^2+h_0^2+\sigma^2-\frac{4}{3}\sigma^2
h_0^2 \frac{1}{h_0^2+\omega^2}
\end{equation}
we can perform Hubbard-Stratonovich transformation and find
representation for partition function with local action (after
redefinition of time scale)
\begin{equation}
S=h\int d\tau \left(
\frac{\dot{x}^2}{2}+\frac{\dot{y}^2}{2}+\frac{\dot{z}^2}{2}
+\left(1+\frac{\sigma^2}{h_0^2}\right)\frac{y^2}{2}+\frac{z^2}{2}+
dy\cos2x+\sqrt{\frac{4}{3}}\frac{\sigma}{h_0} y
z\right)\label{decoupling_random_h}
\end{equation}
\begin{figure}
\psfrag{deldel0}[c][c]{$\frac{\delta\Phi^c}{(\delta\Phi^c)_{reg}}$}
\psfrag{sh0}[c][c]{$\sigma/h_0$}
\hspace{1cm}\includegraphics[width=300pt]{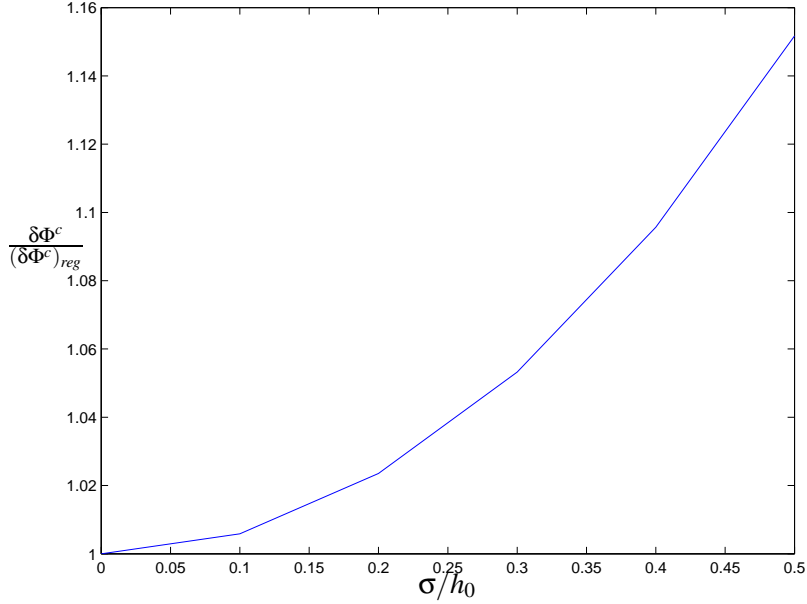}
 \caption{\small Critical deviation from the maximally frustrated point in a chain with disordered flux in rhombi}
\label{random_h}
\end{figure}
Note that our formulation of the problem is a bit specific: we fix relative diversity $\sigma/h_0$
and look for $h_0$ which brings the chain to the point with equal $4e$- and $2e$-supercurrents.

Discussion of the previous section concerning two types of
tunneling trajectories and their connection to components of
supercurrent can be literally applied to action
(\ref{decoupling_random_h}). So we need to estimate  classical
action for $2e$- and $4e$-trajectories. Problem with $\sigma=0$
was analysed in \cite{my}. It was shown that $2e$-supercurrent
dominates at $d\ll 1$ while $4e$-supercurrent --- at $d\gg1$.
Crossover takes place at $d\approx 3.24$. From
(\ref{decoupling_random_h}) one can easily see that corrections to
classical actions due to nonzero $\sigma$ should be of order
$(\sigma/h)^2$. We analyse action (\ref{decoupling_random_h}) numerically.
Results are presented on figure \ref{random_h}.

It may come as a surprise, that with this formulation of the problem critical deviation $\delta\Phi^c$
(determined  through mean flux in rhombi) {\it grows} with $\sigma/h_0$. However, this is quit reasonable since
zero deviation from the maximally frustrated point for {\it one} rhombi in a chain immediately prohibits
$2e$-supercurrent. Allowing for diversity of fluxes in rhombi we allow some of the rhombi to be closer to the
maximally frustrated point than a "mean" rhombi. This fact causes strong suppression of $2e$-supercurrent.

So we conclude that randomness in fluxes is not important  for the pairing effect,
 at least if the standard scatter of these fluxes $\Delta\Phi$
is not too large in comparison with the critical value $\delta\Phi^{c}$ found for  a regular chain,
cf.Eq.(\ref{phi_c_reg}).

\section{Conclusions}
In this paper we provide detailed calculations of superconductive
current in a long frustrated rhombi chain with quenched disorder.
We have considered two types of disorder: random stray charges and random fluxes in rhombi.
We found that small (as compared to the critical deflection $\left(\delta\Phi^c\right)_{reg}$ destroying
$4e$-supercurrent in a clean chain) fluctuations of fluxes piercing rhombi are not that important for the pairing effect.

Main results of our paper concern effect of quenched random stray charges on pairing of Cooper pairs.
For a chain which is relatively far from the maximally frustrated point in the sense $h\gg w$ we managed to calculate
probability to find the system in the regime with dominating $4e$-supercurrent. Stray charges, in principal, may
significantly affect properties of the rhombi chain. In particular, in such a chain $2e$-supercurrent exist even at
the maximally frustrated point. However, as we have demonstrated in Sec. III, at large $E_J/E_C$ and $\Phi_r=\Phi_0/2$
probability to find large $2e$-supercurrent  is small. This result itself is not a great surprise: in a perfectly
classical chain stray charges have no effect at all. More important are two things: i) it is possible to combine
low probability of finding significant $2e$-supercurrent at the maximally frustrated point with exponential
suppression of the supercurrent by quantum fluctuations; ii) in a perfectly classical chain critical deflection
$\delta\Phi^c$ scales with number of rhombi as $1/N$ (this can be easily seen from
eq. (\ref{old E_under_flux_dif_from_pi}) describing energy spectrum of classical rhombi chain) while in a disordered quantum chain this dependence is much
weaker $\delta\Phi^c\sim1/N^{1/6}$, c.f. (\ref{phi_c_random_charges}).

In Sec. IV we have considered modulation of the supercurrent by capacitively coupled gate. In a regular chain, applying
gate voltage one suppresses quantum fluctuations of rhombi and {\it increases} the supercurrent. Its dependence on the
gate voltage is very strong: change of the {\it logarithm} of the supercurrent is linear in $N$. On the contrary, in a chain
with strong random stray charges applying external gate can both {\it increase} or {\it decrease} supercurrent.
Dependence of
the supercurrent on the gate voltage is now much weaker: typical change of the logarithm of the
supercurrent is now proportional to $\sqrt{N}$. Still we see that dependence of the current on gate voltage should be
measurable. Such a dependence is one of the possible ways to demonstrate coherence of quantum phase slips in the chain.

We are grateful to L.~B.~Ioffe and B.~Pannetier for many useful and inspiring
discussions.
This research was supported by  the Program ``Quantum Macrophysics" of
the Russian Academy of Sciences, by the Russian Ministry of Education and Science via the contract
RI-112/001/417  and by Russian Foundation for Basic Research under the
grant No.\ 04-02-16348.
I.V.P. acknowledges financial support from the Dynasty Foundation and Landau-Juelich Scholarship.

\end{document}